\documentclass[12pt,preprint,longnamesfirst]{aastex}

\newcommand{\ie}{\emph{i.e.} }
\newcommand{\ebv}{\ensuremath{\mathrm{E}(\mathrm{B}-\mathrm{V})}}

\slugcomment{Tentatively scheduled for the ApJ 20 August 2005 issue}

\shorttitle{Imaging Polarimetry of Historical Supernov\ae}
\shortauthors{Romaniello et al.}

\begin{document}

\title{VLT-FORS1 Imaging Polarimetry of M83 (NGC~5236)\\I. Search for Light
Echoes from Historical Supernov\ae$^\dag$}
\renewcommand{\thefootnote}{\fnsymbol{footnote}}\footnotetext[2]{Based
on Observations collected at the European Southern Observatory, Chile
(ESO Programme 66.A-0397).}\renewcommand{\thefootnote}{\arabic{footnote}}

\author{Martino Romaniello and Ferdinando Patat}
\affil{European Southern Observatory, Karl-Schwarzschild-Strasse 2,
D-85748 Garching bei M\"unchen, Germany}
\email{mromanie@eso.org, fpatat@eso.org}

\and

\author{Nino Panagia\altaffilmark{1} and William B. Sparks}
\affil{Space Telescope Science Institute, 3700 San Martin Drive, Baltimore,
MD 21218}
\email{panagia@stsci.edu, sparks@stsci.edu}

\and

\author{Roberto Gilmozzi and Jason Spyromilio}
\affil{European Southern Observatory, Paranal Observatory, Chile}
\email{rgilmozz@eso.org, jspyromi@eso.org}

\altaffiltext{1}{On assignment from the Space Telescope Division,
Research and Scientific Support Department of ESA.}

\begin{abstract}
We have used FORS1 at the ESO VLT to search for light echoes in
imaging polarimetry from four historical supernov\ae\ in the face-on
nearby spiral galaxy M83 (NGC~5236).  No echoes were detected around
our targets (SN~1923A, SN~1945B, SN~1950B and SN~1957D). This implies
that the interstellar medium in their environs is rather tenuous
(a few particles per $\mathrm{cm}^{3}$), possibly as a result
of previous supernova explosions that could have cleared the immediate
vicinities of our targets. The merits and limitations of searching for
light echoes in imaging polarimetry are discussed.  From the
photometry of the sources detected at the supernova locations, we
estimate star cluster masses of 720, 400, 300$~M_{\sun}$ for the
cluster progenitors of SN~1957D, SN~1923A, and SN~1950B, respectively,
and an upper limit of few tens of solar masses for SN~1945B.
\end{abstract}


\section{Introduction\label{sec:intro}}
\begin{figure}
  \plotone{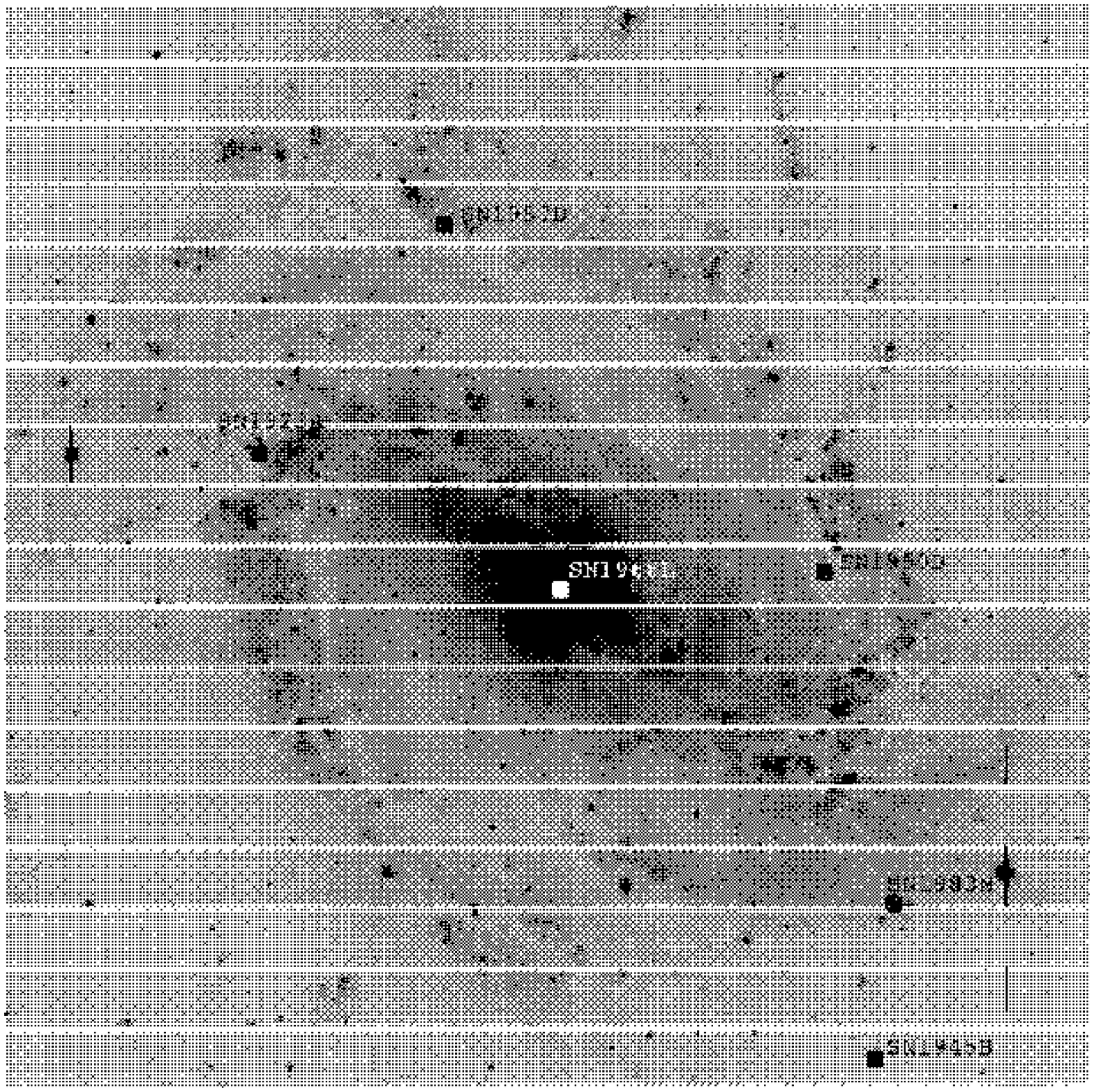}
  \caption{Location of the 6 historical supernov\ae\ known in M83
   superposed to an V-band image of the galaxy.  The image is
   negative, \ie dark objects on a light background. Regrettably
   SN~1983N falls into the gap between two pointings (see text) and
   SN~1968L is lost because it is located on the the bright, heavily
   saturated galactic nucleus. \label{fig:sn_loc}}
\end{figure}

We have undertaken an observational campaign to study the optical
polarization properties of the face-on, nearby spiral galaxy
\objectname{M83} (\objectname{NGC~5236}) from local (\ie few
arcseconds) to galactic (\ie several arcminutes) scales. In this first
paper of a projected series of three, we discuss our search for light
echoes from four out of the six historical supernov\ae\ known in M83,
namely \objectname{SN~1923A}, \objectname{SN~1945B},
\objectname{SN~1950B} and \objectname{SN~1957D}. Their location within
the galaxy is shown in Figure~\ref{fig:sn_loc}, together with that of
\objectname{SN~1983N} and \objectname{SN~1968L}, which we could not
observe because they fall into the gap between two adjacent pointings
and onto the bright, heavily saturated galactic nucleus,
respectively. With the loci of maximum polarization ranging in
diameter from $1.2$ to $3\farcs1$, these supernov\ae\ could have
produced echoes resolvable with ground-based observations under
suitable seeing conditions.

The paper is organized as follows. In section~\ref{sec:echoes} we
briefly summarize the main properties of light echoes and their
connection with the characteristics of the supernova and of the
scattering material. The observations and data reduction are described
in section~\ref{sec:obs} and the resulting polarization maps for the
individual historical supernov\ae\ are presented in
section~\ref{sec:ind}. Finally, the conclusions are drawn in
section~\ref{sec:disc}.

\section{Phenomenology of light echoes\label{sec:echoes}}
Under suitable circumstances after a supernova explodes, due to light
travel time effects, an echo or reflection of light from the explosion
scattered by interstellar matter reaches the observers at a later
date.  This phenomenon is known as a ``light echo''. Of course any
other transient source of light, like a nova, for example, can produce
an echo by the same mechanism. Echoes from supernov\ae, especially type Ia,
are particularly interesting because they can be produced by
dust clouds several hundred parsecs from the supernova, and are, at
least in principle, visible out to distances of a few tens of
megaparsecs from the Sun.  To date, however, only four cases of echoes
from supernov\ae\ at optical wavelengths are known:
\objectname{SN~1991T} \citep[type Ia,][]{sch94,spa99},
\objectname{SN~1998bu} \citep[type Ia,][]{cap01},
\objectname{SN~1987A} \citep[type II,e.g.][]{xu94,spy95} and
\objectname{SN~1993J} \citep[type II,][]{sug02}.  In addition, the
infrared excess observed in some supernov\ae\ has also been
interpreted as due to the presence of a light echo \citep[see,
e.g.,][and references therein]{ger02}. In this paper we will only be
concerned with the optical echoes, whose luminosity is produced by
simple scattering, and not with the infrared ones, which are thought
to originate through thermal emission by dust heated by the supernova
flash \citep{emm88}.

In the single scattering approximation, which we will adopt throughout
this paper, the geometry of an optical echo is very simple \citep[see,
for example,][]{che86,spa94,sug03,pat05}. At any given time the
scattering surface is the iso-delay surface for the light traveling to
the observer, \emph{i.e.} an ellipsoid with the supernova in one focus
and the observer in the other one. In all astrophysical situations,
the distance between the observer and the supernova is much larger
than the one between the supernova and the scattering material and,
for all practical purposes, the ellipsoid can be approximated by a
paraboloid.  The echo, then, crosses the interstellar medium (ISM) out
to large distances from the supernova and can, in principle, be used
as an effective tool to study the ISM properties \citep[spatial
distribution, density, etc; see, e.g.,][and references therein]{xu94}.

Given their simple geometry, \citet{spa94} proposed to use
\emph{imaging polarization observations} of light echoes as a
\emph{purely geometrical} tool to measure distances. As projected on
the sky, light from the center of the echo is backscattered and is not
polarized, light from the outer fringes of the echo is forward
scattered and is weakly polarized, but light initially in the plane of
the sky is scattered at an $90^{\circ}$ angle and is very highly
polarized.  So, when seen through a polarizer, the echo projected on
the sky will have a maximum corresponding to about $90^{\circ}$
scattering. Given the axial symmetry of the system, the locus of
maximum polarization will be a ring in the plane of the sky containing
the supernova (or parts of it, if the scattering material is not
distributed uniformly). Once an echo is identified in polarized light,
measuring the distance to the supernova, and to its parent galaxy, is
straightforward. The \emph{absolute} radius of the ring of maximum
polarization is $r_0=c\ T$, where $T$ is the age of the supernova and
$c$ is the speed of light. If we now call $\phi_0$ the observed
\emph{angular} radius of the ring of maximum polarization, we
immediately get the distance $D$ to the supernova to be $D =
cT/\phi_0$. Under plausible conditions the integrated luminosity of
the echo is expected to be about 10 magnitudes fainter that the
supernova at maximum, so that 8-10 meter class telescopes are needed
to detect these echoes in polarized light beyond the Local Group.

\subsection{Unpolarized light\label{sec:unpol}}
Placing the supernova at the origin of the coordinate system, the
equation of the paraboloidal scattering surface is \citep[see,
e.g.,][]{che86}:

\begin{equation}
  z=\frac{1}{2\ c\ T}\ \left(x^2+y^2\right) - \frac{c\ T}{2}
  \label{eq:par}
\end{equation}
where $x$ and $y$ are the coordinates on the plane of the sky and $z$
is the coordinate perpendicular to it, pointing towards the observer.

Following \citet{spa94}, the surface brightness $F$ as a function of
wavelength of an echo at an angular distance $\phi$ from a supernova of
apparent flux at maximum $S_0$ immersed in a medium of uniform density
is:

\begin{equation}
  F(\lambda)=S_0(\lambda)\ \Phi(\lambda)\ \tau_{\mathrm{echo,sca}}(\lambda)\
    \frac{2}{\phi^2_0+\phi^2}
  \label{eq:F}
\end{equation}
where $\tau_{\mathrm{echo,sca}}$ is the scattering optical depth of
the echo\footnote{This is the optical depth \emph{through} the width
of the expanding echo, not the optical depth \emph{to} the echo itself
due to the material along the line of sight to the observer (see
equation~(\ref{eq:deftau})).} and $\phi_0=cT/D$ ($T$ is the age of the
supernova at the epoch of observations and $D$ is its distance from
the Sun). The ingredients which enter in equation~(\ref{eq:F}) are:

\begin{itemize}
  \item The apparent peak flux {\boldmath$S_0$} of the supernova. The
  peak $V$ magnitudes $m_{\mathrm{max},V}$ of the four historical
  supernov\ae\ we have observed is listed in Table~\ref{tab:sn}
  \citep{bar99}.

  \item The phase function {\boldmath$\Phi$} which appears in
  equation~(\ref{eq:F}) is a function of the scattering angle $\theta$
  and is customarily parametrized in terms of $g(\lambda)$ \citep[$g=0$ for
  isotropic scattering and $g=1$ for pure forward scattering;][]{hen41}
  as:

\begin{equation}
  \Phi(\lambda)=\frac{1-g(\lambda)^2}{4\pi(1+g(\lambda)^2-2g(\lambda)
    \cos\theta)^{3/2}}
  \label{eq:Phi}
\end{equation}

  \item The scattering optical depth
  {\boldmath$\tau_{\mathrm{echo,sca}}$}, which is derived in
  appendix~\ref{sec:tau}, is given by:

\begin{equation}
  \tau_{\mathrm{echo,sca}}(\lambda)= 7.6\times 10^{-5}
  \left(\frac{5.8\times10^{21}}{k}\right) \left(\frac{R(\lambda)}{3.1}\right)
  \left(\frac{\Delta t_{\mathrm{SN}}(\lambda)}{100\mathrm{d}}\right)
  \left(\frac{\omega(\lambda)}{0.6}\right)
  \left(\frac{n_{\mathrm{H}}}{1~\mathrm{cm}^{-3}}\right)
  \left(\frac{n_{\mathrm{O}}}{n_{\mathrm{O,\sun}}}\right)
  \label{eq:tau}
\end{equation}
  where $k=N(H)/\ebv$ is the ratio of total neutral hydrogen surface
  density to color excess at solar metallicity
  \citep[$k=5.8\times10^{21}~\mathrm{cm}^{-2}\,
  \mathrm{mag}^{-1}$,][]{boh78}, $R(\lambda)$ is the ratio of the
  total to selective extinction \citep[in the Milky way on average
  $R(V)=3.1$, e.g.][]{sav79}, $\Delta t_{\mathrm{SN}}(\lambda)$ is the
  duration of the burst of the supernova \citep[$\Delta
  t_{\mathrm{SN}}(V)\simeq100~\mathrm{days}$ for a type~II plateau
  supernova, e.g.][]{pat94}, $\omega(\lambda)$ is the grain albedo
  \citep[$\omega\simeq0.6$ at optical wavelengths, e.g.][]{mat77},
  $n_{\mathrm{H}}$ the hydrogen number density and $n_{\mathrm{O}}$
  the oxygen abundance. We use $n_{\mathrm{O}}=2\ n_{\mathrm{O,\sun}}$
  as determined by \citet{bre02} from several HII regions in M83
  \citep[here we adopt $12+\log(\mathrm{O}/\mathrm{H})_{\sun}=8.69$
  from][] {all01}. In the following, then, we will use a value of
  $\tau_{\mathrm{echo,sca}}(V)=1.5\times 10^{-4}$ as reference for the
  scattering optical depth of supernova echoes in M83. As noticed in
  appendix~\ref{sec:tau}, $R/k$ is the effective cross section per
  hydrogen atom and its value is about
  $5.3\times10^{-22}~\mathrm{cm}^{-2}$ at optical wavelengths
  \citep[see also][]{dra03}.
\end{itemize}

We can now use equation~(\ref{eq:F}) to compute the $V$-band surface
brightness in magnitudes per square arcseconds:

\begin{eqnarray}
  \Sigma_{\mathrm{echo},V}
  & = & m_{\mathrm{max},V}+2.5\lg(\phi_0^2)-2.5\lg\left[\Phi\,
  \tau_{\mathrm{echo,sca}}(V)\, \frac{2}{1+(\phi/\phi_0)^2}\right]\nonumber\\
  & \equiv & m_{\mathrm{max},V}+2.5\lg(\phi_0^2)+\rho(\phi/\phi_0)
  \label{eq:mu}
\end{eqnarray}
where we have introduced the function $\rho(\phi/\phi_0)$ that
contains all of the information about the \emph{shape} of the surface
brightness profile of the echo. This function is plotted in
Figure~\ref{fig:radprof} with a solid line for the case of isotropic
scattering ($g=0$) and with a dashed line for the one in which forward
scattering is favored \citep[$g(V)=0.6$, e.g.][]{dra03}. As discussed
above, a value of $\tau_{\mathrm{echo,sca}}(V)=1.5\times 10^{-4}$ was
used.

\subsection{Polarized light\label{sec:pol}}
The polarization signal expected from a light echo is linear
polarization arranged in a tangential field centered at the supernova
position \citep[e.g.,][]{spa94,pat05}. If there is dust on the plane
of the supernova, the radius of the maximum polarization is $r_0=c\
T$, where $T$ is the age of the supernova at the epoch of the
observations (2001.3, in our case) and $c$ is the speed of light.  At
a distance of $D=4.5$~Mpc \citep[][from Cepheid variables]{thi03} this
translates into an angular radius $\phi_0$ (in arcseconds) given by:

\begin{equation}
  \phi_0[\arcsec]=0.063\ \frac{T[\mathrm{yr}]}{D[\mathrm{Mpc}]}=
             1.4\times10^{-2}\ T[\mathrm{yr}]
  \label{eq:dT}
\end{equation}

The intrinsic degree of polarization is expected to be of the order of
50\% or higher \citep[e.g.,][]{spa94,pat05}. Observationally, it is,
then, diluted because the components of the Stokes vector are combined
vectorially inside each resolution element. Given the image quality in
the final frames of $0\farcs35$ (HWHM) and typical ring radii of
$0.5-1~\arcsec$ from equation~(\ref{eq:dT}), we estimate that the
observed polarization is reduced to about 0.6 times its intrinsic
value, \ie to 30-60\%. Of course, the observed pattern needs not be a
full ring if the dust is not distributed uniformly.

The echo origin of such a structure, if detected, should, then, be
confirmed with spectroscopy. The spectrum of an echo is expected to be
the weighted sum of the spectra of the parent supernova as it evolves
with time, corrected for the scattering extinction efficiency as a
function of wavelength \citep[e.g.,][]{pat05}.

\section{Observations and data reduction\label{sec:obs}}
\begin{figure}
  \plottwo{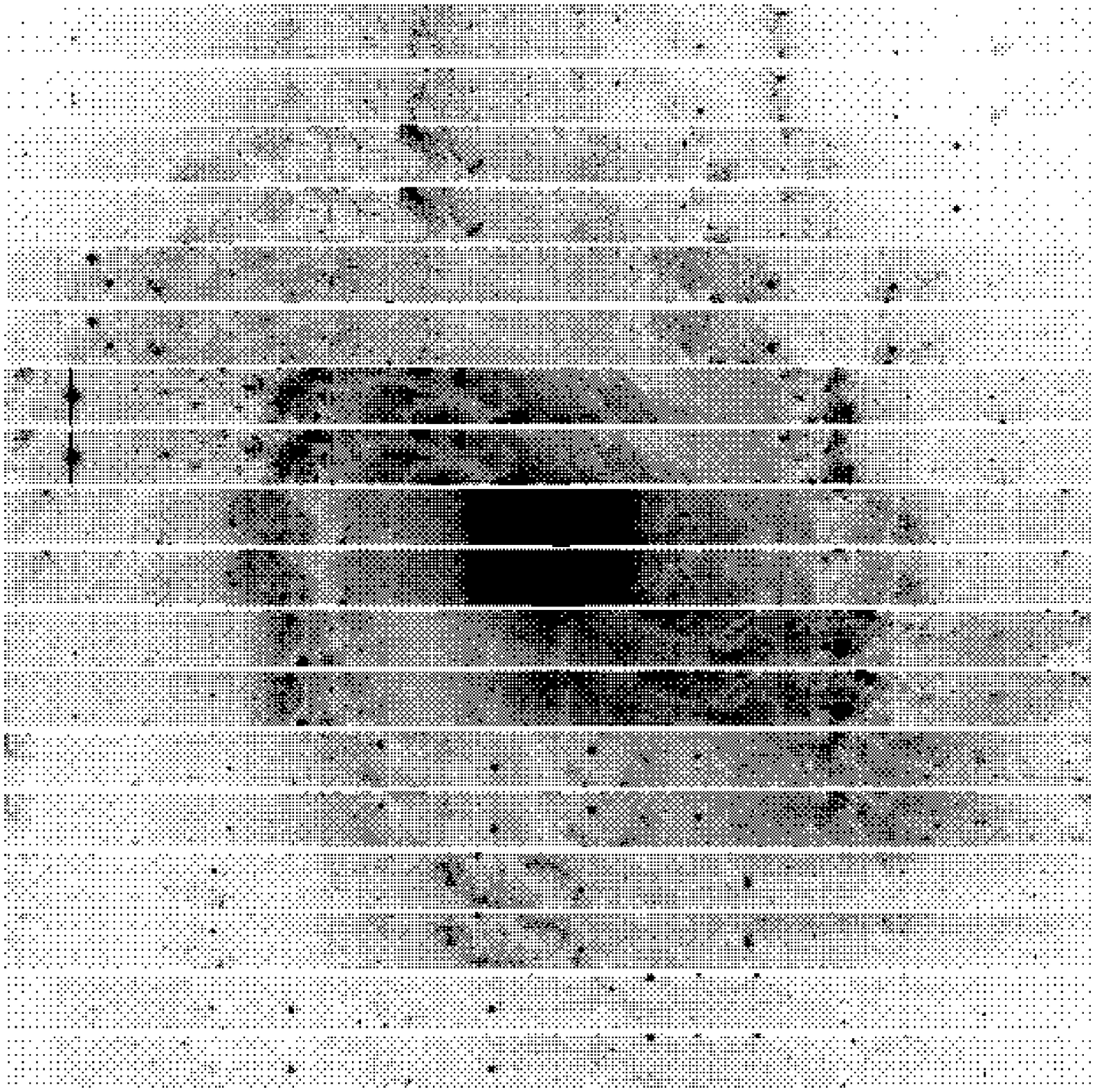}{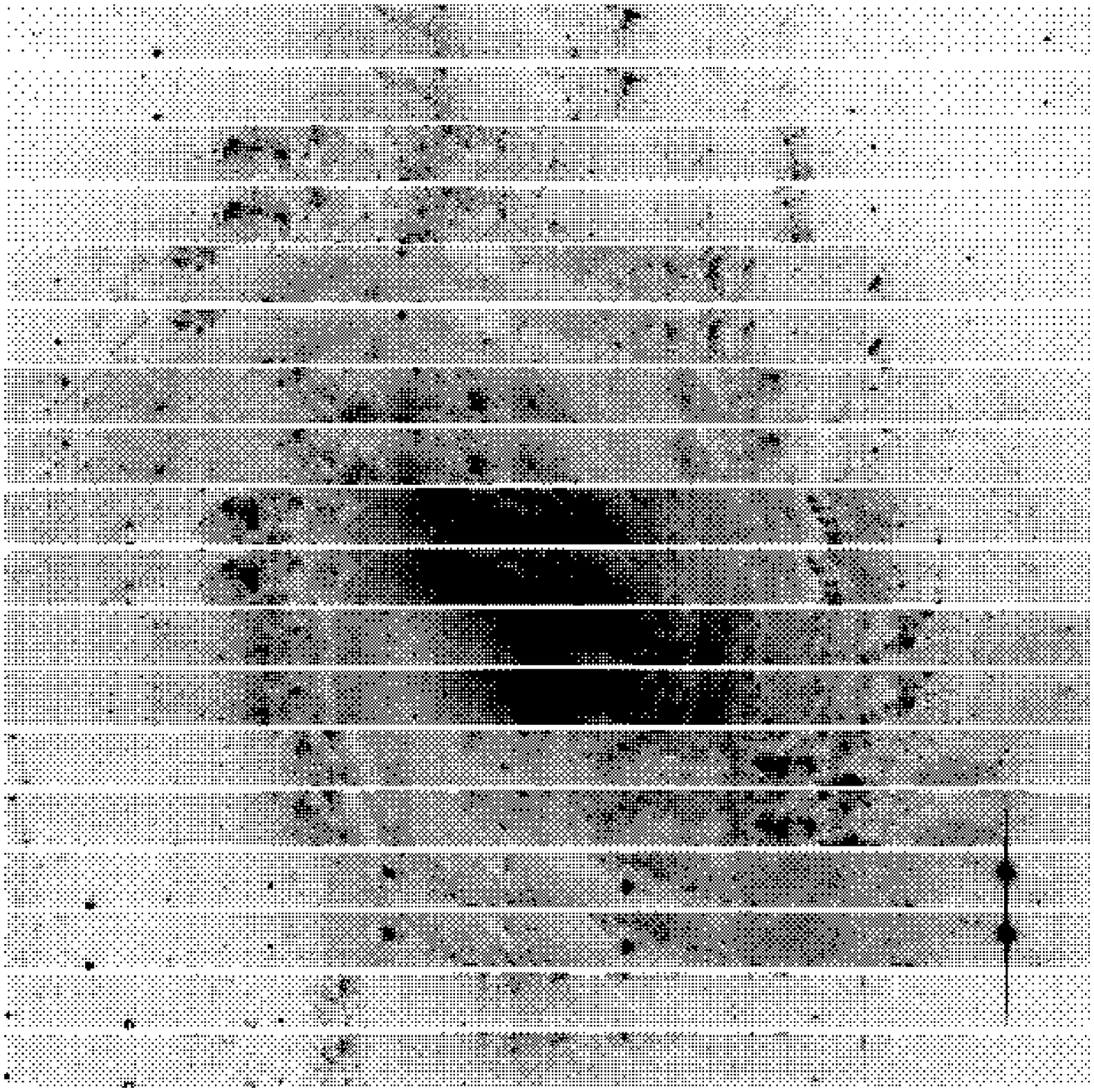}
  \caption{M83 as seen through the polarizing optics of the FORS1
  instrument.  The two pointings we have used to image the whole
  galaxy, offset by $22\arcsec$, are shown in the right and left panel.
  The ordinary and extraordinary beams are clearly visible in both
  images.
  \label{fig:ipol}}
\end{figure}

We have imaged M83 in March 2001 with the \facility{FORS1} instrument
in polarimetric mode at the \facility{ESO Cerro Paranal
Observatory}. The instrument is described in detail in
\citet{sze03}. Polarimetry with FORS1 is performed by inserting in the
light path the combination of a Half Wave Plate and Wollaston
prism. The Half Wave Plate introduces a phase shift in the incoming
light beam, which is equivalent to rotating the instrument as a
whole. The Wollaston prism splits the incoming light in two
orthogonally polarized beams.  A strip mask is inserted in the focal
plane to avoid the overlap of these two beams of polarized light on
the CCD detector plane. As a result, every exposure covers half of the
field of view in two mutually orthogonal linear polarization
directions, usually referred to as ordinary and extraordinary beams.
These, then, need to be combined to derive the Stokes parameters and,
since they were taken simultaneously, the effects of variations in the
the Earth's atmosphere transparency are canceled \citep[see,
e.g.,][]{app67,sca83}. In order to cover the entire galaxy, we have
taken two sets of exposures, shifted by one slit throw
($22\arcsec$). The coordinates and rotator angle were chosen so as to
place as many historical supernov\ae\ as possible (five out of six) in
the same pointing. Regrettably, SN~1968L is projected on the bright
galactic nucleus, which is heavily saturated in our images, and could
not be studied. Two examples of the resulting images are shown in
Figure~\ref{fig:ipol}.

Every pointing was observed in the V band using 4 Half Wave Plate
angles spaced by $22.5\degr$, \ie 8 polarization directions. The total
observing time at each Half Wave Plate position was 1.9 hours, split
into 15 minute exposures for accurate cosmic ray removal. The image
quality on the final stacked image is $0\farcs7$ FWHM. A redundancy of
the polarization directions is recommended to mitigate instrumental
imperfections (improper flat-fielding, non-ideal Wollaston
transmission, etc.). The astrometric solution was reconstructed with a
few stars from the \dataset{GSC2} present in the FORS1 field of view.

After de-biasing and flat-fielding with unpolarized sky flat field
frames\footnote{see also
\url{http://www.eso.org/instruments/fors1/pola.html.}}, the images for
the 4 Half Wave Plate positions were carefully aligned and combined to
derive the Stokes parameters and, ultimately, the polarization degree
and angle with their associated errors following the procedures as
discussed in \citet{pr04}.

The value of the local background for the ordinary and extraordinary
beams around the position of each supernova was estimated as the
minimum intensity in a $6\arcsec$ square box \citep[130~pc at
4.5~Mpc,][]{thi03} centered on the supernova itself. The background
shows some spatial structure even on these relatively small
scales. The adoption of a single value for the background is justified
by noticing that different prescriptions on how to estimate it
(minimum value, median value, mode, a simple 2D model with a low-order
polynomial) all lead to the same results to within the random
observational uncertainties. It is still possible, though, that,
because of the complex spatial distribution of the diffuse emission,
we are not subtracting the ``true'' value of the background. This could
introduce a systematic bias in the inferred degree of polarization,
which is rather hard to estimate. On the other hand, the polarization
angle measured with dual-beam instruments like FORS1 only depends on
the ratio of normalized flux differences in the ordinary and
extraordinary beams \citep[e.g.,][]{pr04}. As such, it is to the first
order independent of the exact value used to estimate and subtract the
background.  The polarization pattern is, then, determined rather
robustly and can be used as a good diagnostics for the presence of a
light echo (tangential polarization).

Quantitatively, the typical measured background around the position of
the historical supernov\ae\ is $20.9~\mathrm{mag}/\sq\arcsec$. In the
exposure time we have used, 1.9 hours for each of the $N=4$ Half Wave
Plate angles needed to carefully derive the polarization degree ($P$)
and angle ($\chi$), it leads to a signal-to-noise ratio (SNR) per
pixel of 4 for a source with $26~\mathrm{mag}/\sq\arcsec$. In turn,
this corresponds to an uncertainty on $P$ and $\chi$ of \citep[see,
e.g.,][]{pr04}:

\begin{equation}
  \sigma_\mathrm{P}=\left[\sqrt\frac{N}{2}\ \mathrm{SNR}\right]^{-1}
     \simeq15\%
  \label{eq:sp}
\end{equation}

\begin{equation}
  \sigma_\chi=\left[2\sqrt\frac{N}{2}\ \mathrm{SNR}\ P\right]^{-1}\simeq
     12\degr
  \label{eq:sc}
\end{equation}
where in the last equation we have used $P=40\%$.

Finally, all of the images taken with different Half Wave Plate angles
can be combined together to reach a higher signal-to-noise ratio for
detection of unpolarized light. Thus, the $4\sigma$ limit in
unpolarized light is $26.7~\mathrm{mag}/\sq\arcsec$ or, conversely, an
unpolarized emission of $26~\mathrm{mag}/\sq\arcsec$ is detectable at
an $8\sigma$ level in our data.

\section{Polarization around the individual supernov\ae\label{sec:ind}}
The polarization maps in the neighborhood of the four historical
supernov\ae\ we have studied in M83 are shown in
Figures~\ref{fig:sn57d} through \ref{fig:sn50b}. The location where
the maximum of the polarization is expected according to
equation~(\ref{eq:dT}) is marked with a circle. Let us recall here
that we do not have data for the other two historical supernov\ae\ in
M83, SN~1983N and SN~1968L, because the first one happened to be
located in our images in a gap between two pointings and the latter
one is projected on the the bright, heavily saturated galactic
nucleus.

\subsection{SN~1957D\label{sec:57d}}
The generally accepted value for the $V$ magnitude at maximum of
SN~1957D is $m_{\mathrm{max},V}=15$ \citep[e.g.][]{bar99}. At a
distance of 4.5~Mpc, \ie a distance modulus of 28.3, it translates
into an intrinsic magnitude of $-13.3$, which places SN~1957D at the
very faint end of the type~II supernov\ae\ luminosity function, even
fainter then SN~1987A \citep[for reference, a ``regular'' type~II
peaks at $M_{\mathrm{max},V}\simeq-16$ for
$H_0=75\, \mathrm{km}~\mathrm{s}^{-1}~\mathrm{Mpc}^{-1}$;][]{pat94}. It
is possible, then, that SN~1957D was observed after maximum and that
it was, in reality, brighter than $15^{\mathrm{th}}$ magnitude. If
this were the case, then, the echo would be intrinsically brighter
than reported in Table~\ref{tab:sn}. For lack of better evidence,
though, in the following we will adopt $m_{\mathrm{max},V}=15$.

\begin{figure}
  \plottwo{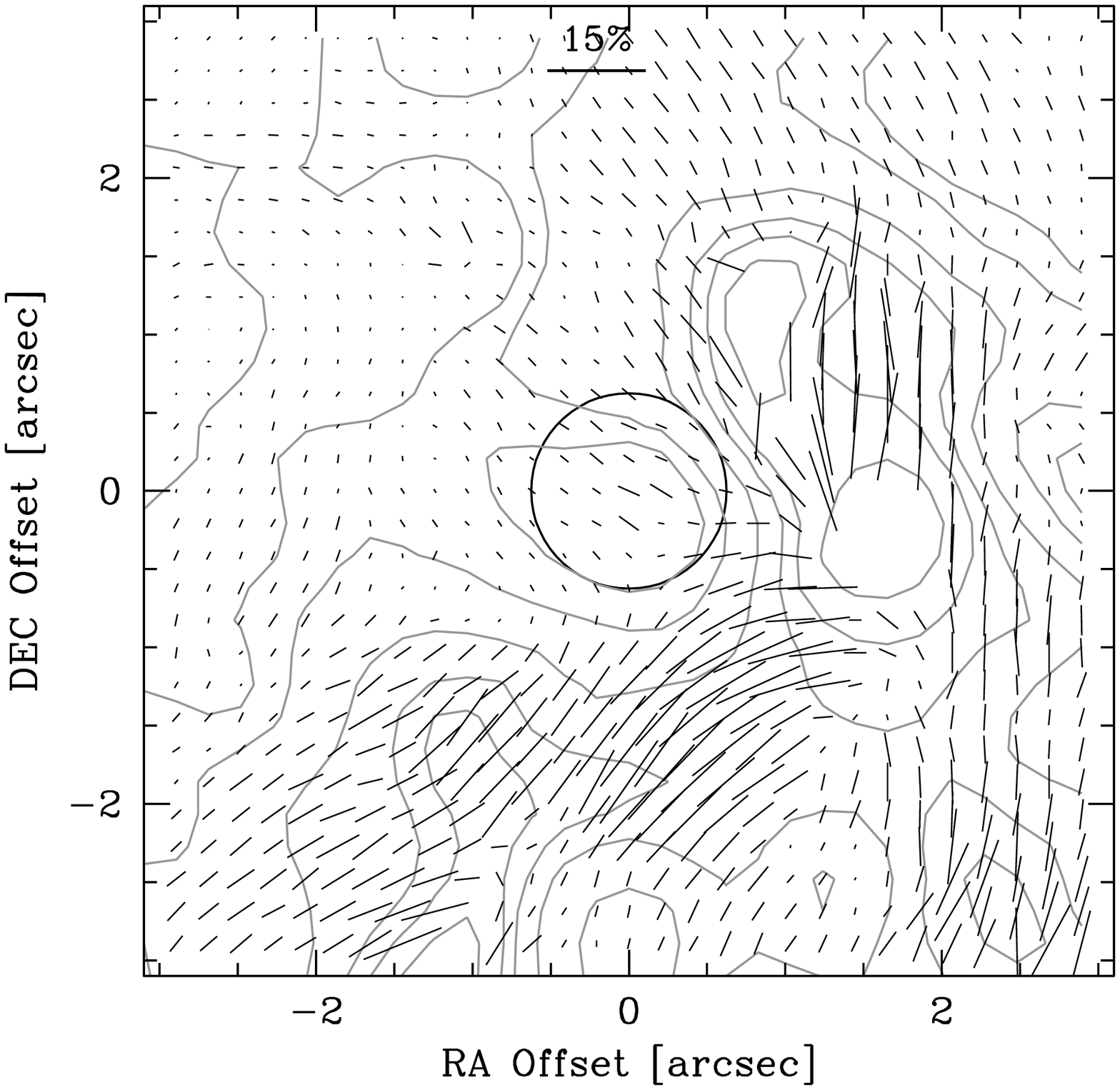}{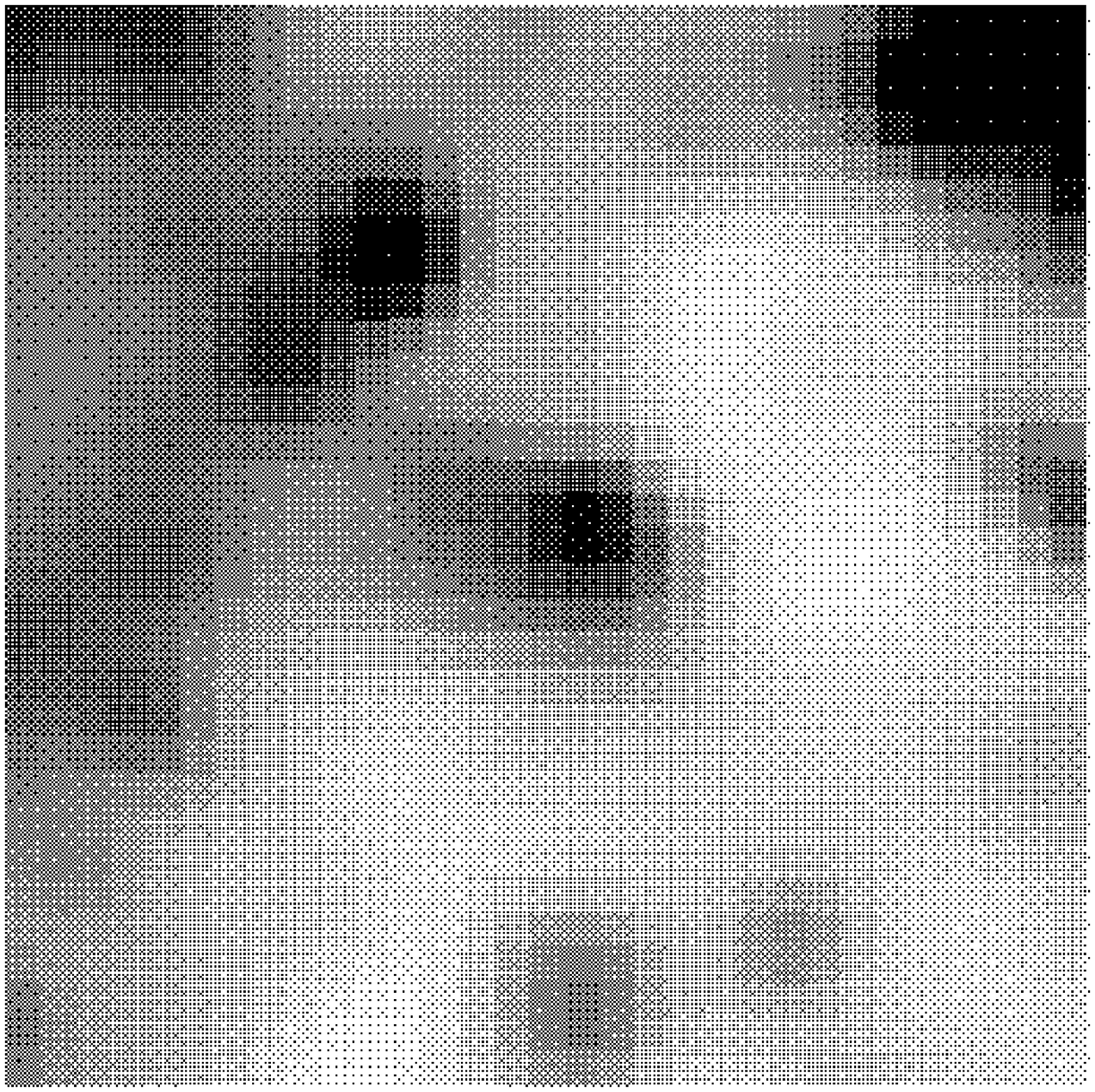}
  \caption{Polarization (\emph{left panel}) and intensity maps
  (\emph{right panel}) within a box of $6\arcsec$ (13~parsecs)
  centered on SN~1957D. North is up and East to the left. The circle
  in the left panel marks the expected location of the polarization
  maximum given a distance of 4.5~Mpc to M83 \citep{thi03} and under
  the assumption of dust on the plane of the supernova (see
  equation~(\ref{eq:dT})). The image in the right panel is negative,
  \ie dark objects on a light background.\label{fig:sn57d}}
\end{figure}

\begin{figure}
  \plottwo{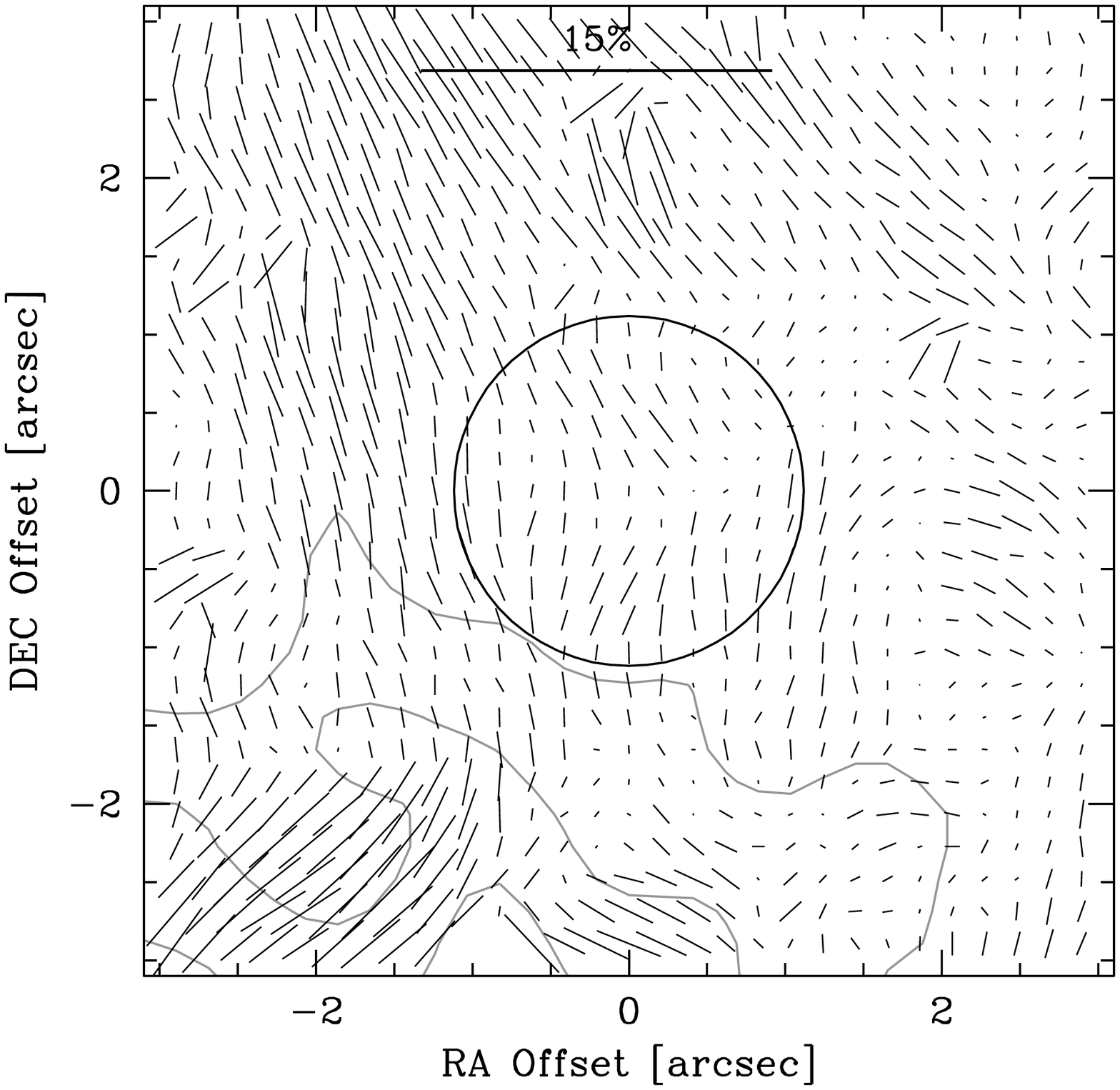}{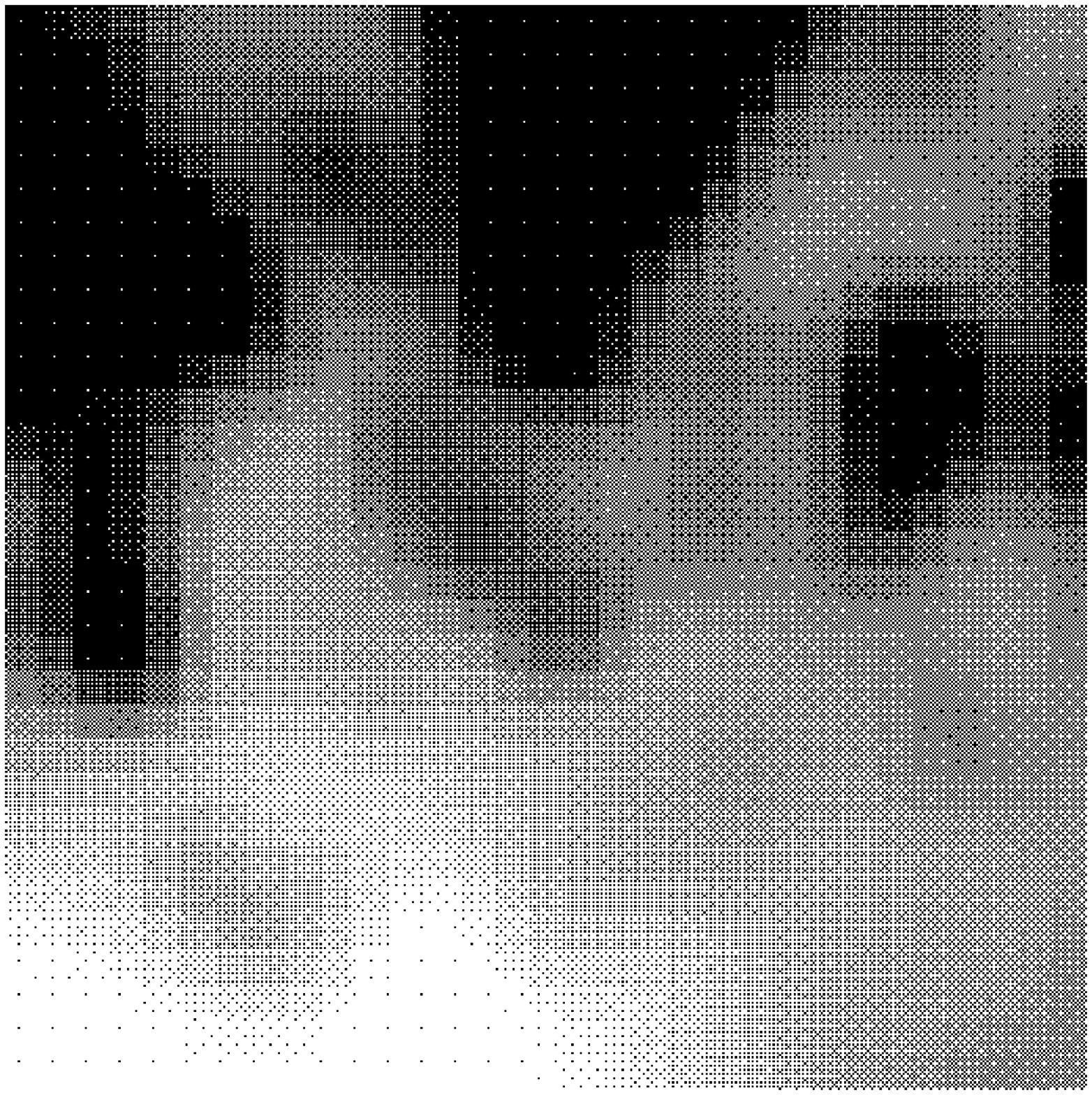}
  \caption{Same as Figure~\ref{fig:sn57d}, but for SN~1923A.\label{fig:sn23a}}
\end{figure}

\begin{figure}
  \plottwo{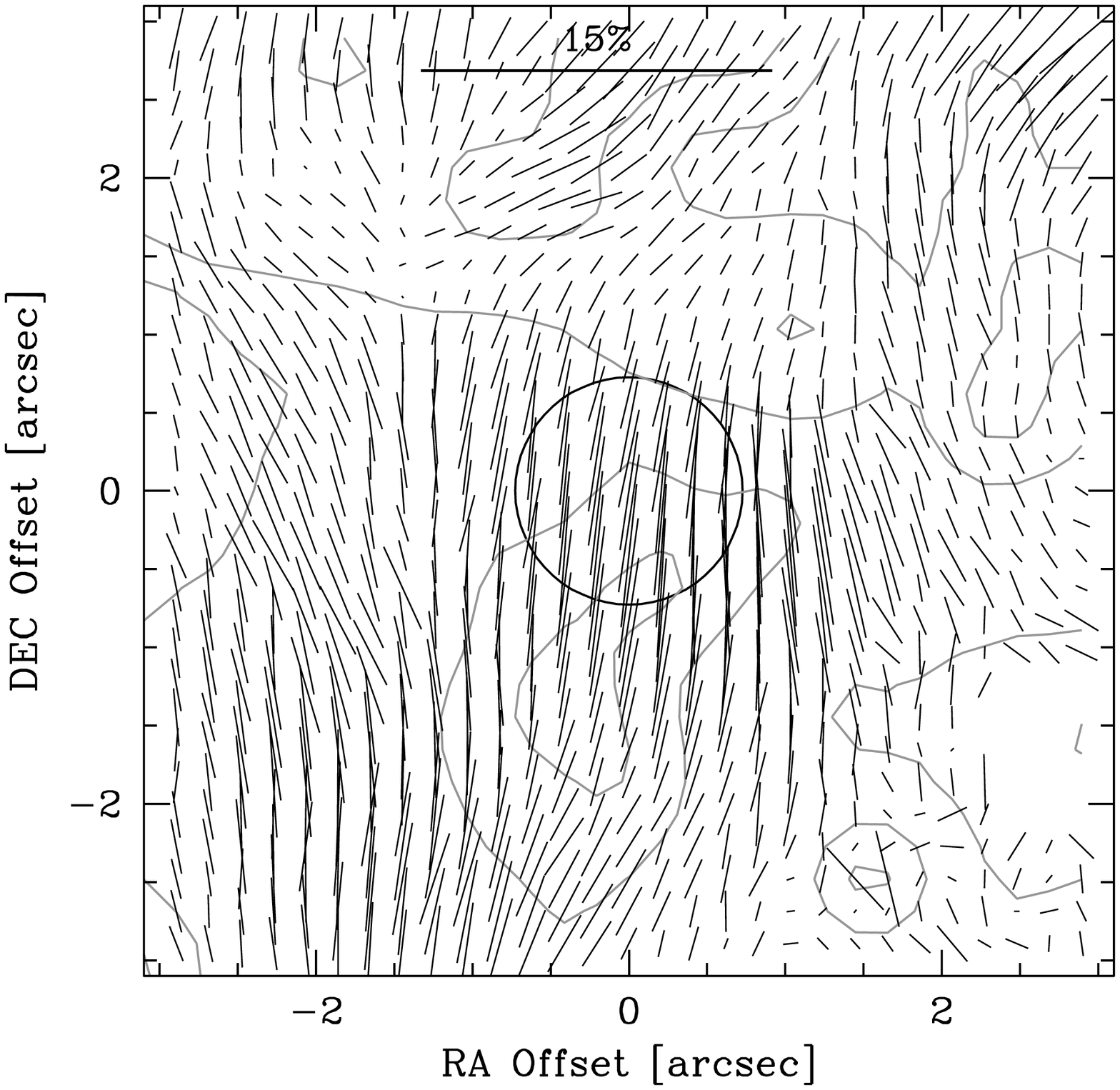}{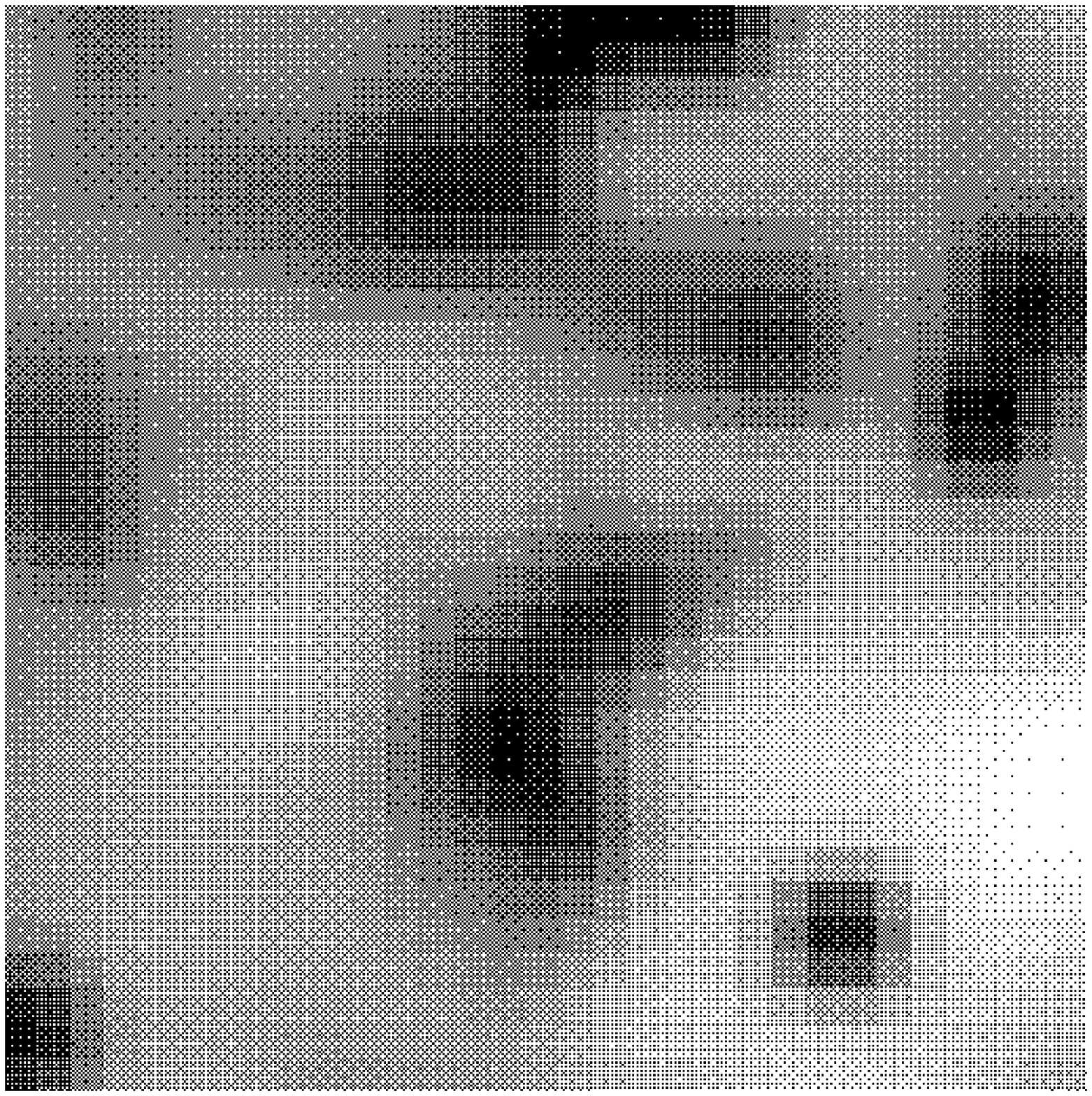}
  \caption{Same as Figure~\ref{fig:sn57d}, but for SN~1950B.\label{fig:sn50b}}
\end{figure}

The polarization around SN~1957D is displayed in
Figure~\ref{fig:sn57d}, together with an image of the same region of
$6\arcsec\ (130~\mathrm{pc})$ on the side. Clear structures with
polarization degrees of several percents are detected. However, it is
not possible to attribute any of these structures to the light from
the supernova being scattered by interstellar clouds, because the
polarization vectors are mostly oriented \emph{radially} relative to
the position of the supernova itself. On the contrary, as discussed in
section~\ref{sec:pol}, the expected pattern for an echo would be
\emph{tangential}. These polarized emissions are most likely
associated with the large-scale polarization of M83, which follows the
spatial distribution of dust along the spiral arms. Its study will be
the subject of a forthcoming paper (Romaniello et al, in preparation).
The simplest explanation for the non-detection of an echo is that
there is not enough dust in the vicinity of the supernova to produce
one \citep[in fact,][find a rather modest value of reddening to
SN~1957D itself, $E(B-V)\simeq0.1$]{lon92}. We will come back to this
in section~\ref{sec:disc}.

A rather strong unpolarized emission is also seen at the position of
the supernova (Figure~\ref{fig:sn57d}, right panel), which is likely
to be associated with the supernova remnant itself. Its total
magnitude, as measured on our images, is $V=22.7$. The detection of
this compact emission is highly significant, with a signal-to-noise
ratio of about 100. The most recent spectroscopy at the position of
SN~1957D by \citet{tur89} and \citet{lon92} suggests that about 50\%
of the $V$-band flux from the central source is emitted in the very
broad [OIII] $\lambda\lambda 4959, 5007$ emission ($\mathrm{FWHM}\sim
2700~\mathrm{km}~\mathrm{s}^{-1}$, with the red wing of the line
extending $4500~\mathrm{km}~\mathrm{s}^{-1}$ beyond the zero velocity
position). \citet{tur89} measured a flux of $\sim
2\times10^{-15}\mathrm{ergs}~\mathrm{cm}^{-2}~\mathrm{s}^{-1}$ for the
[OIII] blend and detect several other lines (MgI $\lambda 4571$, [OI]
$\lambda 6300,63$ and H$\alpha$+[NII] $\lambda\lambda 6548,83$ [OII]
$\lambda\lambda 7319,30$) and interpret the spectrum as that of a
supernova remnant, similar to \object{Cas~A} in the Galaxy and
\object{N132D} in the Large Magellanic Cloud \citep[see
also][]{lon92}.  In our images the emission is spatially resolved,
with a radius of $0\farcs55$, or 12~pc at 4.5~Mpc \citep{thi03}. This
size is appreciably larger than expected for a supernova remnant
\citep[for reference, Cas~A is 2~pc in radius at an age of about 300
years,][]{ray84}, thus indicating that the remnant itself is embedded
in an HII region and/or a group of stars. Actually, \citet{lon92}
detected a narrow H$\alpha$ emission component which is indicative of
an HII region, and \citet{cow94} infer the presence of a substantially
emitting HII region from the observed flattening of the spectral index
of the radio flux from SN~1957D.

We have run a simple simulation to evaluate the probability that the
spatial correspondence between the location of SN~1957D and the HII
region is due to a chance alignment. By generating random locations
for the supernova over the observed spatial distribution of the
sources in its vicinity we estimate the probability of a chance
alignment to be less than 20\%.

Conservatively, attributing 50\% of the $V$ band luminosity to the
stellar component and adopting a distance modulus of 28.3, the
underlying stellar cluster would have an absolute magnitude
$M_V=-4.8-A_V\simeq-5.2$ for an average visual extinction of
0.35~magnitudes \citep{lon92}. Using Starburst99 models \citep{lei99}
for a metallicity of twice solar, as appropriate at the galactocentric
distance of SN~1957D \citep{bre02}, and a Salpeter IMF within the
interval $1-100~M_{\sun}$, such a magnitude corresponds to a stellar
cluster of initial mass of about $720~M_{\sun}$ if SN~1957D was the
``median" supernova to explode in the cluster, \ie when the cluster
was at an age of about 16~Myrs \citep{pen82}. Had the cluster had a
younger age, a lower cluster mass would be inferred, and viceversa.
With an initial mass of $720~M_{\sun}$ and an age of 16~Myrs, the
cluster should have produced about 7 supernov\ae, and correspondingly
released a total $\sim 10^{52}~\mathrm{ergs}$ in kinetic energy. This
energy input is enough to sweep away the ISM from the immediate
vicinity of the supernova, hence justifying our non-detection of light
echoes in its immediate proximity.

Finally, let us note that \citet{bof99} have indicated SN~1957D as a
possible candidate for hosting an echo based on the observed blue
$\bv$ and $\vr$ colors of a compact ``feature'' detected
$6\arcsec~\mathrm{W}$ and $1\farcs9~\mathrm{N}$ of the nominal
position of the supernova, \ie outside of the box in the right-hand
panel of Figure~\ref{fig:sn57d}. As such, the object they identify is
unlikely to be related either to the supernova, or to a possible echo.

\subsection{SN~1923A, SN~1945B and SN~1950B\label{sec:nodet}}
\begin{figure}
  \plottwo{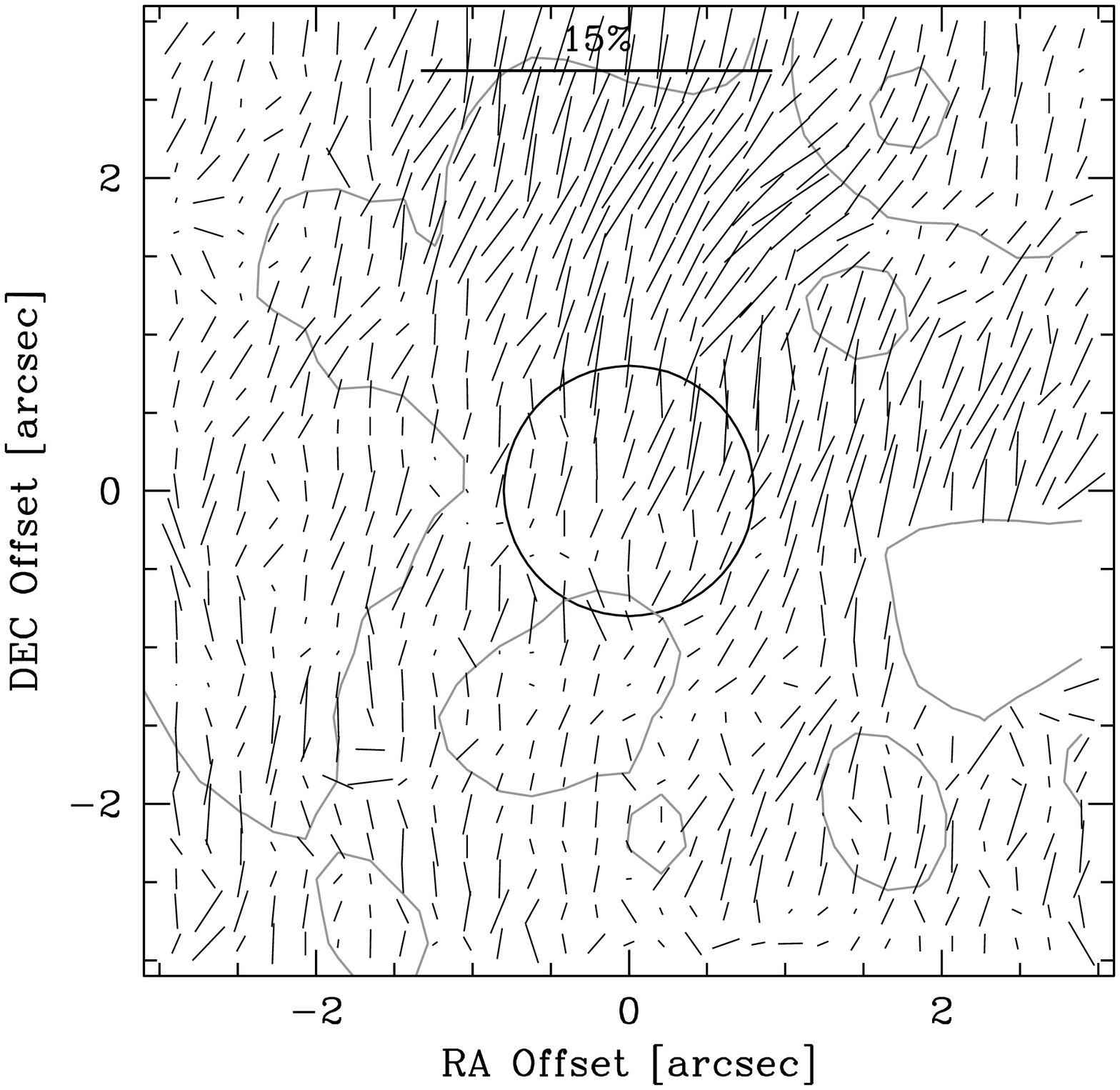}{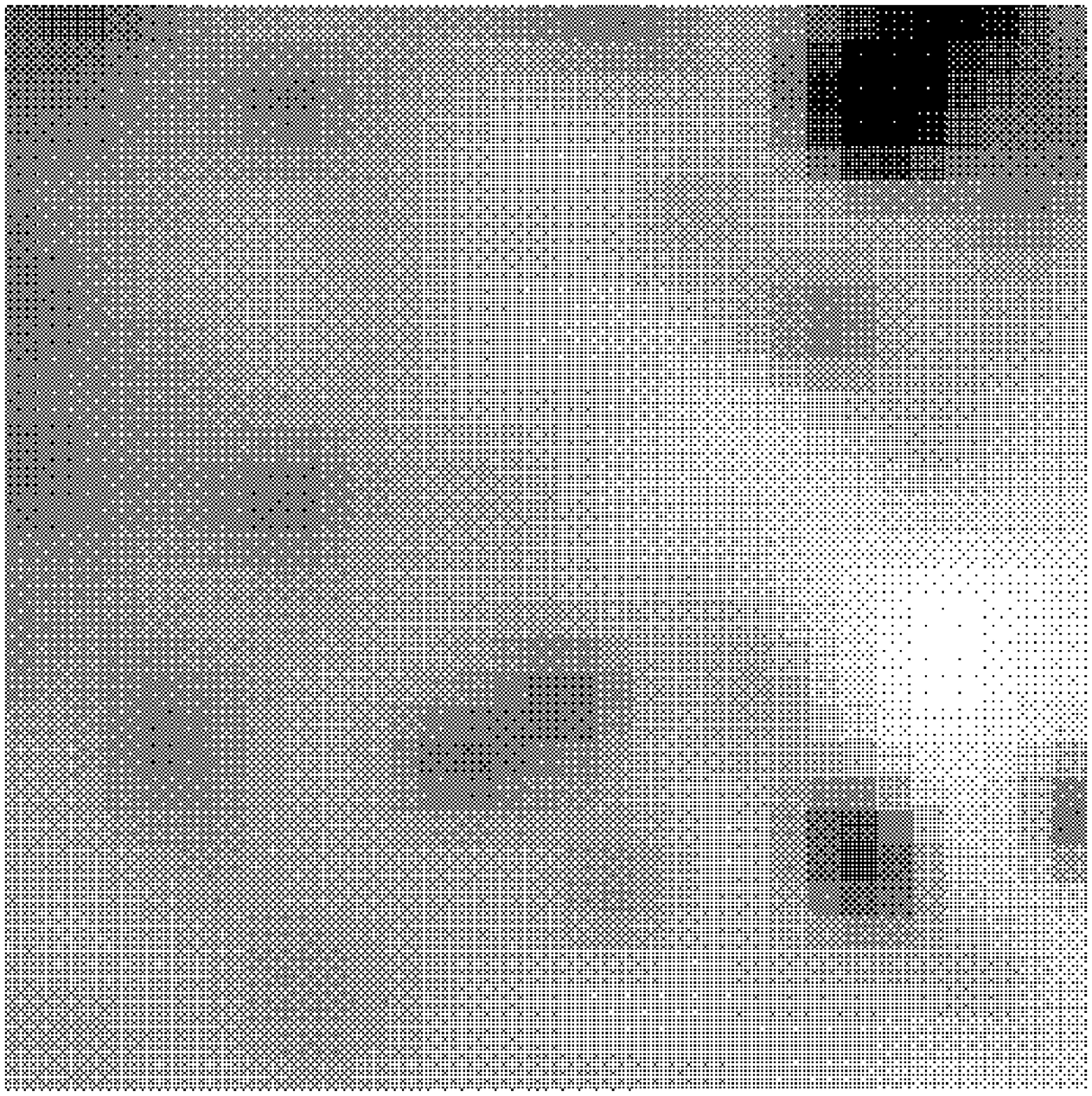}
  \caption{Same as Figure~\ref{fig:sn57d}, but for SN~1945B.\label{fig:sn45b}}
\end{figure}

In Figures~\ref{fig:sn23a} through \ref{fig:sn50b} we show the
polarization and intensity maps for the surroundings of SN~1923A,
SN~1945B and SN~1950B. Again, polarization degrees of several percents
are detected, but there is no sign of a coherent circular structure
for any of these three supernov\ae.  As we noticed in
section~\ref{sec:57d}, the simplest explanation for the non-detection
of an echo is that there is not enough dust in the vicinity of the
supernova to produce one. This goes in the same direction, but on
larger spatial scales, as for the results of \citet{bla04}, who did
not find evidences of remnants for any of these three supernov\ae.

Let us now use the $V$ magnitude measured in our images at the nominal
position of the historical supernov\ae\ to estimate the mass of the
stellar population that gave birth to SN~1923A, SN~1945B and SN~1950B,
again under the assumption that each of them was the "median" one to
explode among its population. As discussed above, this assumption
implies an age of about 16~Myrs for the stars. The emission at the
position of SN~1923A is measured to have $V=23.3$ and that at the
position of SN~1950B $V=23.7$. Fluxes were measured with aperture
photometry and both detections are highly significant, at the 60 and
$40\sigma$ level, respectively.

As in the case of SN~1957D, we have run a simple simulation to
evaluate the probability that the spatial correspondence between the
locations of SN~1923A and SN~1950B and the optical emissions detected
in our images is due to a chance alignment. By generating random
locations for the supernov\ae\ over the observed spatial distribution
of the sources in their vicinity we have estimated the probability of
a chance alignment to be of the order of 30\%. This is slightly larger
than in the case of SN~1957D because the remnant itself was not
detected neither for SN~1923A nor SN~1950B and, hence, we have allowed
for an error-box of $1\arcsec$ on the published positions of the
supernov\ae\ \citep{bar99}.

Using the stellar synthesis models of \citet{lei99} as for the case of
SN~1957D, these magnitudes imply initial star cluster masses of about
400 and $300~M_{\sun}$ for SN~1923A and SN~1950B, respectively. For
SN~1945B, on the other hand, we could only measure an upper limit for
the emission at the location of the supernova (see
Figure~\ref{fig:sn45b}, right panel) of $V=27$ at $4\sigma$, implying
that the parent cluster contained only a few tens of solar masses.

\section{Discussion and conclusions\label{sec:disc}}
In this paper we have presented very deep VLT-FORS1 imaging
polarimetry in the V band aimed at detecting and studying the light
echoes from four of the six historical supernov\ae\ known in the
face-on spiral galaxy M83. We took advantage of the multiplex
capabilities afforded by the FORS1 instrument in its polarimetric mode
by observing the targets simultaneously. The exposure time for each of
the four Half Wave Plate angles needed to recover the polarization
degree and angle was 1.9 hours.

M83 is a well suited target to search for light echoes because, being an
actively star-forming galaxy  \citep[e.g.][and references therein]{bla04},
it has hosted 6 known historical supernov\ae. Moreover, its proximity to
the Sun \citep[4.5~Mpc,][]{thi03}, makes it possible to resolve the echoes
in seeing-limited observations with
8-10 meter class ground-based telescopes. In fact, the final image
quality on our stacked image is $0\farcs7$ (FWHM), which would allow
the detection of echoes, if present, as close as a $\phi_0=cT/D$ to
the parent supernova (see Table~\ref{tab:sn}).  In spite of this and
of our very deep exposures (7.6 hours for every pointing), we did not
detect the characteristic tangential polarization pattern around any
of the four supernov\ae\ we have studied (SN~1923A, SN~1945 and
SN~1950B, and SN~1957D). If echos were produced around the targets,
then, they have to be fainter than our detection threshold in
polarized light of $26~\mathrm{mag}/\sq\arcsec$ (see
section~\ref{sec:obs}).

\begin{figure}
\plotone{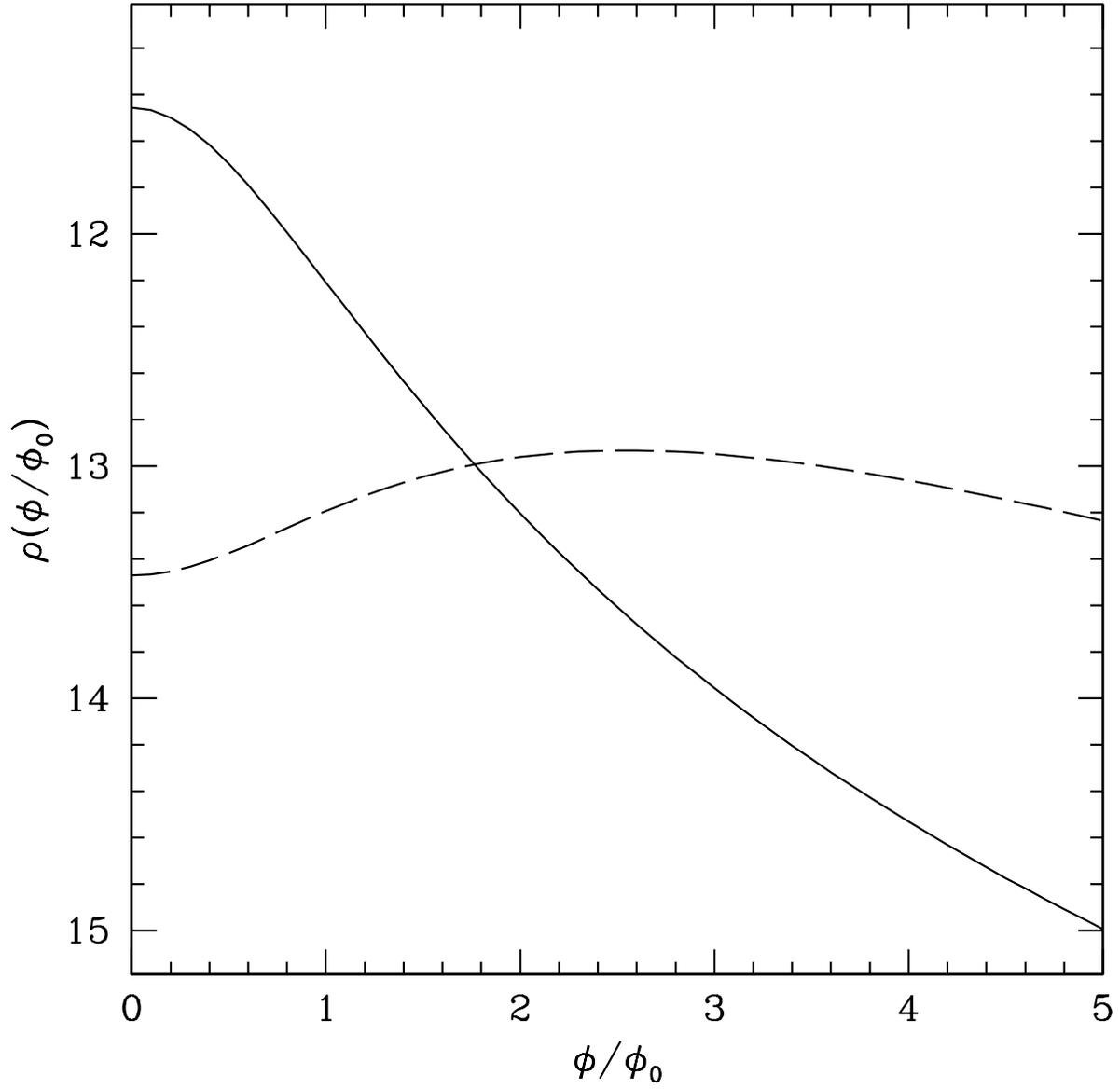}
\caption{Expected radial profile of the light echoes for uniformly
distributed dust from equation~(\ref{eq:mu}) for isotropic ($g=0$,
solid line) and forward-favored scattering ($g=0.6$, dashed line). As
discussed in the text, a value of $\tau_{\mathrm{echo,sca}}=1.5\times
10^{-4}$ was used.
\label{fig:radprof}}
\end{figure}

At this stage, then, we can only place an upper limit to the
scattering optical depth in the vicinities of our target supernov\ae,
which can, then, be translated into an upper limit on the density of
the interstellar material.  By combining equation~(\ref{eq:mu}) with
the values of $m_{\mathrm{max,}V}+2.5\lg(\phi_0^2)$ listed in
Table~\ref{tab:sn} and the echo radial profile of
Figure~\ref{fig:radprof}, we see that the typical surface brightness
expected for the echoes at a radial distance of
$\phi=1\mathrm{-}2~\phi_0$ is of about $27~\mathrm{mag}/\sq\arcsec$
for the reference value of the scattering optical depth of
$\tau_\mathrm{echo,sca}(V)=1.5\times10^{-4}$.  Of course, the higher
$\tau_\mathrm{echo,sca}$, the brighter the echo. The fact that no
echoes were detected above our threshold of
$26~\mathrm{mag}/\sq\arcsec$, then, implies that the actual value of
$\tau_\mathrm{echo,sca}(V)$ in the vicinity of the historical
supernov\ae\ has to be lower than about $4\times10^{-4}$.

According to equation~(\ref{eq:tau}) a higher value of the optical
depth $\tau_{\mathrm{echo,sca}}$ compared to the reference value can
be due to a combination of a higher effective cross section $R/k$,
and/or a longer duration of the burst of the supernova ($\Delta
t_{\mathrm{SN}}$), and/or higher albedo and/or high density of the
interstellar medium ($n_{\mathrm{H}}$). Of these four quantities, we
consider the latter one as the most likely to determine a variation of
the value of the optical depth in different environments. In fact, a
cross section of about $5.3\times10^{-22}~\mathrm{cm}^{-2}$ at optical
wavelengths is appropriate for typical dust \citep[e.g.][and
references therein]{dra03}, supernov\ae\ rarely have bursts lasting
longer than 100 days \citep[e.g.][]{pat94} and the albedo in the
optical is rather firmly established to be $\omega\simeq0.6$
\citep[e.g.][]{mat77}. We, thus, conclude, that the non-detection of
light echoes implies a rather tenuous interstellar medium around the
four historical supernov\ae, with a hydrogen number density lower than
about $n_{\mathrm{H}}=2.5~\mathrm{cm}^{-3}$.

Equation~(\ref{eq:tau}) makes it easy to adapt the result to different
characteristics of the dust (\ie the effective cross section $R/k$,
the albedo $\omega$), of the parent supernova (\ie the duration of the
burst $\Delta t_\mathrm{SN}$), or the metallicity of the interstellar
medium ($n_\mathrm{O}$). In particular, we have used a typical
Galactic value of $R=3.1$ for the ratio of the total to selective
extinction \citep[e.g.][]{sav79,dra03}, which seems to be appropriate
for M83 as a whole \citep{boi05}. $R$ is observed to vary between
about 2 and 5.5 along different lines of sight in the Milky Way and
other galaxies \citep[and references therein]{dra03}, which would
cause the inferred value of $n_\mathrm{H}$ to increase or,
respectively, decrease by the same, rather mild, factor.

The energy input from previous supernov\ae\ in these clusters could be
responsible for sweeping away the ISM from the immediate vicinities of
our targets. If these low values are common, detecting and
characterizing light echoes, even from objects as intrinsically
luminous as supernov\ae\ in nearby galaxies, would be an almost
prohibitive task in terms of telescope time.

From the photometry of the sources detected at the supernova
locations, we estimate star cluster masses of 720, 400, 300$~M_{\sun}$
for the cluster progenitors of SN~1957D, SN~1923A, and SN~1950B,
respectively, and an upper limit of few tens of solar masses for
SN~1945B.

To conclude, we would like to emphasize that identifying light echoes
in imaging polarimetry is advantageous because they should produce a
clear signature in the form of a tangential polarized structure around
the parent supernova. On the other hand, though, detailed dust
modeling indicate that forward scattering is a better description of
reality, rather than the simpler isotropic case \citep[e.g.][and
references therein]{dra03}. As shown in Figure~\ref{fig:radprof} light
echoes from a uniform distribution of dust are expected to be much
more extended in the case of forward scattering (dashed line) than in
the one of isotropic scattering (solid line), in which they have a
characteristic size of $\phi_o=cT/D$. Disentangling such an extended
source superposed on a complex galactic background is an extremely
challenging task (see also the discussion in
section~\ref{sec:obs}). Ultimately, background subtraction could be
the main limiting factor in the detection of echoes, both in
unpolarized and polarized light. Therefore, searches should be focused
on the outskirts of galaxies, where the background is smoother, hence
easier to evaluate and subtract.

\appendix
\section{Derivation of equation~(\ref{eq:tau})\label{sec:tau}}
In this appendix we derive the expression in equation~(\ref{eq:tau})
for the scattering optical depth through the expanding flash
$\tau_{\mathrm{echo,sca}}$:

\begin{equation}
  \tau_{\mathrm{echo,sca}}(\lambda)=c\, \Delta
  t_{\mathrm{SN}}(\lambda)\, n_d\, \sigma_{\mathrm{sca}}(\lambda)
  \label{eq:deftau}
\end{equation}
where $c$ is the speed of light, $\Delta t_{\mathrm{SN}}(\lambda)$ is
the duration of the burst of the supernova at wavelength $\lambda$,
$n_d$ the space density of the dust scattering particles and
$\sigma_{\mathrm{sca}}(\lambda)$ is the scattering cross section. This
latter quantity is related to the extinction coefficient $A(\lambda)$
through the albedo $\omega(\lambda)$:

\begin{equation}
  A(\lambda)=2.5\log(e)\, \tau_{\mathrm{ext}}(\lambda)=2.5\log(e)\,
  N_d\, \sigma_{\mathrm{ext}}(\lambda)=1.086\, N_d\,
  \frac{\sigma_{\mathrm{sca}}(\lambda)}{\omega(\lambda)}
  \label{eq:Alam}
\end{equation}
where $\tau_{\mathrm{ext}}$ and $\sigma_{\mathrm{ext}}$ are the
extinction optical depth and cross section, respectively, and $N_d$ is
the column density of the scattering dust.

Let us now introduce the ratio of total neutral hydrogen density to
color excess at solar metallicity $k=N(H)/\ebv$, and allow it to
scale linearly with the oxygen abundance $n_{\mathrm{O}}$
\citep{jam02}:

\begin{equation}
  N(H)=k\ \ebv\, \frac{n_{\mathrm{O}}}{n_{\mathrm{O,\sun}}} =k\,
  \frac{A(\lambda)}{R(\lambda)}\,
  \frac{n_{\mathrm{O}}}{n_{\mathrm{O,\sun}}} =1.086\,
  \frac{k}{R(\lambda)}
  \frac{\sigma_{\mathrm{sca}}N_d}{\omega(\lambda)}\,
  \frac{n_{\mathrm{O,\sun}}}{n_{\mathrm{O}}}
  \label{eq:NH}
\end{equation}
where in the last passage we have used equation~(\ref{eq:Alam}).

Combining equations~(\ref{eq:deftau}) and (\ref{eq:NH}) we get to the
desired expression for $\tau_{\mathrm{echo,sca}}$ as a function of
wavelength:

\begin{eqnarray}
  \tau_{\mathrm{echo,sca}}(\lambda) & = & \frac{c}{1.086}\,
  \frac{R(\lambda)}{k}\, \Delta t_{\mathrm{SN}}(\lambda)\, \omega(\lambda)\,
  n_{\mathrm{H}}\,
  \frac{n_{\mathrm{O}}}{n_{\mathrm{O,\sun}}}\nonumber\\ & = &7.6\times
  10^{-5} \left(\frac{5.8\times10^{21}}{k}\right)
  \left(\frac{R(\lambda)}{3.1}\right) \left(\frac{\Delta
  t_{\mathrm{SN}}(\lambda)}{100\mathrm{d}}\right)
  \left(\frac{\omega(\lambda)}{0.6}\right)
  \left(\frac{n_{\mathrm{H}}}{1~\mathrm{cm}^{-3}}\right)
  \left(\frac{n_{\mathrm{O}}}{n_{\mathrm{O,\sun}}}\right)
  \label{eq:atau}
\end{eqnarray}
with $k=5.8\times10^{21}~\mathrm{cm}^{-2}\, \mathrm{mag}^{-1}$ from
\citet{boh78}, $R(V)=3.1$ from \citet{sav79}, $\Delta
t_{\mathrm{SN}}(V)\simeq100~\mathrm{days}$ for plateau type~II
supernov\ae\ from \citet{pat94} and $\omega\simeq0.6$ at optical
wavelengths from \citet{mat77}. The effective cross section per
hydrogen atom $R/k$, then, is about $5.3\times10^{-22}~\mathrm{cm}^2$
at optical wavelengths.

\acknowledgments
We would like to thank the ESO personnel for successfully carrying out
the very demanding Service Mode observations described here: the final
image quality of $0\farcs7$ in the V band over 17 hours of total
execution time is quite a treat! Simone Bianchi, Palle M{\o}ller,
Thomas Szeifert and Jeremy Walsh are gratefully acknowledged for many
stimulating and fruitful discussions. We wish to thank an anonymous
referee whose insightful comments helped us to avoid unwarranted
conclusions about possible detections of light echoes in M83.

\clearpage

\clearpage

\clearpage

\clearpage

\clearpage

\clearpage

\clearpage

\clearpage

\clearpage

\begin{table*}[!ht]
\begin{center}
\begin{tabular}{*{5}{c}}
Object & $m_{\mathrm{max,}V}$ &  $cT$ &  $\phi_0$ & $m_{\mathrm{max,}V}+2.5\lg(\phi_0^2)$ \\
       & [mag]  & [lyrs] & [arcsec] & [$\mathrm{mag}/\sq\arcsec$] \\ \tableline
  SN~1923A   &    14   & 77.8 & 1.1 & 14.2 \\
  SN~1945B   &    13.9 & 55.7 & 0.8 & 13.4 \\
  SN~1950B   &    14.5 & 51.0 & 0.7 & 13.8 \\
  SN~1957D   &    15   & 43.4 & 0.6 & 13.9 \\
\end{tabular}
\end{center}
\caption{Basic properties of the 4 historical supernov\ae\ we have
observed in M83 and their expected light echoes: apparent $V$-band
magnitude of the supernov\ae\ at maximum
\citep[$m_{\mathrm{max,V}}$,][]{bar99}, linear and angular radius of
the polarization maximum of the echo ($cT$ and $\phi_0=cT/D$,
respectively; see equation~(\ref{eq:dT})) and normalization of the
echo surface brightness profile
($m_{\mathrm{max,}V}+2.5\lg(\phi_0^2)$; see equation~(\ref{eq:mu})).
A Cepheid-based distance to M83 of 4.5~Mpc from \citet{thi03} was used
throughout this paper.}
\label{tab:sn}
\end{table*}


\begin{thebibliography}{}
\bibitem[Allende Prieto et al(2001)Allende Prieto, Lambert \& Asplund]{all01}
   Allende Prieto, C., Lambert, D.L., and Asplund, M. 2001, \apj, 556, L63
\bibitem[Appenzeller(1967)]{app67} Appenzeller, I. 1967, \pasp, 79, 136
\bibitem[Barbon et al(1999)]{bar99} Barbon, R., Buond\'i, V., Cappellaro, E.,
   and Turatto, M. 1999, \aaps, 139, 531
\bibitem[Blair \& Long(2004)]{bla04} Blair, W.P., and Long, K.S. 2004,
  \apjs, 155, 101
\bibitem[Boffi et al(1999)Boffi, Sparks \& Macchetto]{bof99} Boffi, F.R.,
   Sparks, W.B., and Macchetto, F.D. 1999, \aaps, 138, 253
\bibitem[Biossier at al(2005)]{boi05} Boissier, S., Gil de Paz, A.,
  Madore, B.F. et al 2005, \apj, 619, L83
\bibitem[Bohlin et al(1978)Bohlin, Savage \& Drake]{boh78} Bohlin, R.C.,
   Savage, B.D., and J.F. Drake 1978, \apj, 224, 132
\bibitem[Bresolin \& Kennicutt(2002)]{bre02} Bresolin, F., and Kennicutt,
   R.C. 2002, \apj, 572, 838
\bibitem[Cappellaro et al(2001)]{cap01} Cappellaro, E., Patat, F., Mazzali,
   P.A., Benetti, S., Danziger, J.I., Pastorello, A., Rizzi, L., Salvo, M.,
   and Turatto, M. 2001, \apj, 459, 215
\bibitem[Chevalier(1986)]{che86} Chevalier, R.A. 1986, \apj, 308, 225
\bibitem[Cowan et al(1994)Cowan, Roberts \& Branch]{cow94} Cowan, J.J.,
   Roberts, D.A., and Branch, D. 1994, \apj, 434, 128
\bibitem[Draine(2003)]{dra03} Draine, B.T. 2003, \araa, 41, 241
\bibitem[Gerardy et al(2002)]{ger02} Gerardy, C.L., Fesen, R.A., Nomoto, K.,
   Garnavich, P.M., Jha, S., Challis, P.M., Kirshner, R.P., H\"oflich, P.,
   and Wheeler, J.C. 2002, \apj, 575, 1007
\bibitem[Emmering \& Chevalier(1988)]{emm88} Emmering, R.T., and
   Chevalier, R.A. 1988, \aj, 95, 152
\bibitem[Henyey \& Greenstein(1941)]{hen41} Henyey, L.C., and Greenstein, J.L.
   1941, \apj, 93, 70
\bibitem[James et al(2002)]{jam02} James A., Dunne, L., Eales, S., and
   Edmunds, M.G. 2002 \mnras, 335, 753
\bibitem[Leitherer et al(1999)]{lei99} Leitherer, C., Schaerer, D.,
   Goldader, J.D., Delgado, R.M. Gonz\'alez, Robert, C., Kune, D.F.,
   de Mello, D.F., Devost, D., Heckman, T.M. 1999, \apjs, 123, 3
\bibitem[Long et al(1992)Long, Winkler \& Blair]{lon92} Long, K.S., Winkler,
   P.F., and Blair, W.P. 1992, \apj, 395, 632
\bibitem[Mathis et al(1977)Mathis, Rumpl \& Nordsieck]{mat77} Mathis, J.S.,
   Rumpl, W., and Nordsieck, K.H. 1977, \apj, 217, 425
\bibitem[Patat(2005)]{pat05} Patat, F. 2005, \mnras, 357, 1161
\bibitem[Patat et al(1994)]{pat94} Patat, F., Barbon, R., Cappellaro,
   E., and Turatto, M. 1994, \aap, 282, 731
\bibitem[Patat \& Romaniello(2005)]{pr04} Patat, F., and Romaniello, M.
   2004, \pasp, in press
\bibitem[Pennington et al(1982)Pennington, Talbot \& Dufour]{pen82}
   Pennington, R.L., Talbot, R.J., Jr., and Dufour, R.J. 1982, \aj, 87, 1538.
\bibitem[Raymond(1984)]{ray84} Raymond, J.C. 1984, \araa, 22, 75
\bibitem[Savage \& Mathis(1979)]{sav79} Savage, B.D., and Mathis, J.S.
   1979, \araa, 17, 73
\bibitem[Scarrott et al(1983)]{sca83} Scarrott, S.M., Warren-Smith, R.F.,
   Pallister, W.S., Axon, D.J., and Bingham, R.G. 1983, \mnras, 204, 1163
\bibitem[Schmidt et al(1994)]{sch94} Schmidt, B.P., Kirshner, R.P.,
   Leibundgut, B., Wells, L.A., Porter, A.C., Ruiz-Lapuente, P., Challis, P.,
   and Filippenko, A.V. 1994, \apj, 439, L19
\bibitem[Sparks(1994)]{spa94} Sparks, W.B. 1994, \apj, 433, 19
\bibitem[Sparks et al(1999)]{spa99} Sparks, W.B., Macchetto, F.D.,
   Panagia, N., Boffi, F., Branch, D., Hazen, M.L., and Della Valle,
   M. 1999, \apj, 523, 585
\bibitem[Spyromilio et al(1995)]{spy95} Spyromilio, J., Malin, D.F.,
   Allen, D.A., Steer, C.J., and Couch, W.J. 1995, \mnras, 274, 256
\bibitem[Sugerman \& Crotts(2002)]{sug02} Sugerman, B.E.K, and Crotts, A.P.S.
   2002, \apj, 581, L97
\bibitem[Sugerman(2003)]{sug03} Sugerman, B.E.K 2003, \apj, 126, 1939
\bibitem[Szeifert \& Jehin(2003)]{sze03} Szeifert, T., and Jehin, E.
   2003, FORS 1+2 User Manual,\\
   \url{http://www.eso.org/instruments/fors/userman}
\bibitem[Thim et al(2003)]{thi03} Thim, F., Tammann, G.A., Saha, A.,
   Dolphin, A., Sandage, A., Tolstoy, E., and Labhardt, L. 2003,
   \apj, 590, 256
\bibitem[Turatto et al(1989) Turatto, Cappellaro \& Danziger]{tur89}
   Turatto, M., Cappellaro, E., and Danziger, J. 1989, The Messenger,
   56, 36
\bibitem[Xu et al(1994)Xu, Crotts \& Kunkel]{xu94} Xu, J., Crotts, A.P.S.,
   and Kunkel, W.E. 1994, \apj, 435, 274
\end{thebibliography}
\end{document}